\documentclass[usegraphicx]{mn2e}
\usepackage{times}
\usepackage{amssymb}
\newcommand{\msun}{$\mathrm{M_{\sun}}$}

\newcommand{\xpn}[2]{$#1 \times 10^{#2}$}
\newcommand{\rate}{$\mathrm{M_{\sun}} \, \mathrm{yr}^{-1}$}

\title[{CXOU} J121538.2+361921: a double neutron star in the making?]
      {{CXOU} J121538.2+361921 in the galaxy {NGC} 4214: a double neutron star in the making?}

\author[Dewi]
       {J. D. M. Dewi\thanks{email: jasinta@ast.cam.ac.uk}\\
        Institute of Astronomy, University of Cambridge,
        Madingley Road, Cambridge CB3 0HA, UK}
\date{Accepted . Received ; in original form }

\pagerange{\pageref{firstpage}--\pageref{lastpage}}
\pubyear{}

\begin{document} 

\maketitle 

\label{firstpage}

\begin{abstract}
CXOU~J121538.2+361921 is the brightest X-ray source in the galaxy NGC~4214, with an X-ray luminosity of up to \xpn{0.7}{39}~$\mathrm{erg \, s^{-1}}$. The observed periodicity of 3.62~hr is interpreted as the orbital period of the system. It has been suggested that the system is a low-mass helium star with a lower-mass compact companion. If this idea is correct, then CXOU~J121538.2+361921 will evolve into a double neutron star, a binary consisting of a radio pulsar and another neutron star. In this study we investigate further this possibility. We find that the X-ray luminosity is consistent with super-Eddington accretion in a helium star-neutron star binary. The binary is in a state of mass transfer phase which is initiated when the helium-star donor is on the helium shell burning stage. A donor star with a current mass in the range of around 2.2 -- 3.6~\msun\ is required to explain the observed orbital period. Helium stars in this mass range are massive enough to collapse in a supernova explosion, making CXOU~J121538.2+361921 the immediate progenitor of a double neutron star.
\end{abstract}


\section{Introduction}
\label{ngc4214:sec:intro}

Double neutron stars (DNSs) are binary systems consisting of a recycled radio pulsar and a usually undetected neutron star (an exception to this is the double pulsar J0737--3039 in which both neutron stars are observed as pulsars; Lyne et al. 2004). Since the first discovery of B1913+16 (Hulse \& Taylor 1975), DNSs have acted as an excellent test bed for the theory of general relativity. The merger of two neutron stars due to gravitational waves -- a phenomenon observable by gravitational wave detectors such as the Laser Interferometer Gravitational Wave Observatory (LIGO), Virgo and the Laser Interferometer Space Antenna (LISA) -- is thought to be the generator of short-duration gamma-ray bursts.

The standard evolutionary scenario for the formation of DNSs involves a Be/X-ray binary stage, a system consisting of a rapidly rotating Be star and an X-ray emitting neutron star (see, e.g.\ Bhattacharya \& van den Heuvel 1991). The Be/X-ray binary experiences a runaway mass-transfer phase leading to a common-envelope and spiral-in phase. The outcome of this phase is the helium core of the Be star and the neutron star in a close orbit. Provided that the helium star is massive enough to collapse and that the supernova kick velocity does not disrupt the binary, the system evolves into a DNS. 

An alternative scenario called the double-core channel, originally proposed by Brown (1995) and developed further by Dewi, Podsiadlowski \& Sena (2006), avoids the Be/X-ray binary stage. In this channel, DNSs originate from binary systems of almost-equal mass which undergo a contact and spiral-in phase, producing a CO core from the primary and a helium core from the secondary. After the CO core collapses into a neutron star, the helium star may transfer matter to the neutron star before it eventually goes supernova. Therefore, the standard and double-core scenarios both share the same latest stage of evolution, i.e.\ the helium star-neutron star (HeS-NS) binary phase.

While the total number of observed Be/X-ray binaries is growing to more than 100 systems both in the Galaxy and the Magellanic Clouds (Liu, van Paradijs \& van den Heuvel 2000; 2005), the only candidate for a HeS-NS binary to date is Cyg~X-3, an X-ray binary with an orbital period of 4.8~hr. The nature of the components of Cyg ~X-3 is still in debate. Although the donor is clearly a hydrogen-depleted star (van Kerkwijk et al. 1992; 1996; Fender, Hanson \& Pooley 1999), the mass -- and hence the fate -- of the donor is not well known. The nature of the compact object, i.e.\ whether it is a neutron star or a black hole, is also still in question. Based on a population synthesis study, Lommen et al. (2005) proposed that the system could consist of either a massive Wolf-Rayet star 
 and  a black hole, or a low-mass helium star of $\lesssim$ 1.5~\msun\ with a neutron-star companion. Only in the latter case can Cyg X-3 act as the progenitor of a DNS. The recent discovery of a 3.62-hr period X-ray binary CXOU~J121538.2+361921 (Ghosh et al. 2006) provides an interesting possibility of another plausible progenitor of a DNS system.

\section{CXOU J121538.2+361921 as the immediate progenitor of a double neutron star} 
\label{ngc4214:sec:progenitor}

\subsection{The system CXOU J121538.2+361921}
\label{ngc4214:subsec:source}

CXOU~J121538.2+361921 is the brightest X-ray source in the dwarf starburst galaxy NGC~4214. It shows a 3.62-hr periodicity -- which is interpreted as the orbital period of the system -- and has an X-ray luminosity of up to \xpn{0.7}{39}~$\mathrm{erg \, s^{-1}}$ (Ghosh et al. 2006). 

A comparison between the mass-radius relation applied to the system and that of hydrogen-rich and helium stars reveals that the donor of CXOU~J121538.2+361921 is either a low-mass ($\sim$ 0.4~\msun) hydrogen-burning star or a helium-burning star and that a mass ratio, $q = M_{\mathrm{donor}} / M_{\mathrm{accretor}}$, larger than 1 is necessary (Ghosh et al. 2006). This minimum $q=1$ is required for an inclination angle of 90$^{o}$ and a minimum eclipse half angle of 22$^{o}$; for a lower inclination or higher minimum eclipse half angle, the minimum mass ratio will be larger. Several possibilities have been put forward to explain the nature of this system, i.e.\ a low-mass X-ray binary, a foreground cataclysmic variable (CV) of AM Her type, or a compact object with a helium-star donor. The first option is considered to be unlikely since the mass-transfer rate from a low-mass star, typically of the order of $10^{-10} - 10^{-9}$~\rate, is too low to produce the observed X-ray luminosity. The lack of a companion down to a visual magnitude of 22.6 indicates that a CV is not probable. For the third alternative, a helium star in the range of $\sim$ 2 -- 23~\msun\ was proposed. 

Fig.~4 in Ghosh et al. (2006) shows that massive helium stars ($\gtrsim$ 12~\msun) can only be the possible donor if the system has a mass ratio larger than unity, is in Roche-lobe overflow (RLOF), and the helium star is not evolved. However, the lack of apparent stellar wind, which is expected from helium stars more massive than $\sim$ 3.5~\msun\ (Pols \& Dewi 2002), suggests that this possibility is excluded. The idea is also rejected on evolutionary grounds as RLOF in systems with massive helium stars on the main sequence is dynamically unstable (Dewi et al. 2002). If the donor is on the terminal age helium-star main sequence, a lower mass is required, i.e.\ $\gtrsim$ 2.2~\msun\ for $q\geq1$. Consequently, RLOF initiated during helium shell burning and beyond results in lower-mass helium stars, and/or lower mass ratios. The low donor mass and a minimum mass ratio of 1 further imply that a neutron star or stellar-mass black-hole companion is allowed.

In this study we will investigate further the likelihood of CXOU~J121538.2+361921 as an exclusively HeS-NS binary, and hence as the immediate progenitor of a DNS.

\subsection{The evolution of helium star -- neutron star binaries}
\label{ngc4214:subsec:hesns}

The X-ray luminosity, $L_{\mathrm{X}}$, of CXOU~J121538.2+361921 is as high as \xpn{0.7}{39}~$\mathrm{erg \, s^{-1}}$. Here $L_{\mathrm{X}}$ can be expressed as $L_{\mathrm{X}} = G M \dot{M} / R$, where $M$ and $R$ are the mass and radius of the compact object, $\dot{M}$ is the mass-accretion rate, and $G$ the gravitational constant. Applying the canonical values for a neutron star, $M = 1.4$~\msun\ and $R = 10$~km, we obtain $\dot{M} = $ \xpn{6}{-8}~\rate, i.e.\ twice the Eddington mass-accretion rate for helium on to a neutron star. The typical mass-transfer rate in HeS-NS binaries is $10^{-5} - 10^{-4}$~\rate\ (Dewi et al. 2002). Hence the suggestion that the system is a HeS-NS is very plausible.

The most recent detailed, extensive studies of the evolution of HeS-NS binaries have been carried out by, e.g., Dewi et al. (2002), Dewi \& Pols (2003) and Ivanova et al. (2003). These studies investigated a wide parameter space in helium-star masses and orbital periods. During the mass-transfer phase, it was assumed that the neutron star accretes matter up to its Eddington limit, i.e.\ \xpn{3}{-8}~\rate, and the rest of the transferred matter is lost from the system, carrying with it the specific angular momentum of the neutron star. The orbital evolution includes the loss of angular momentum due to helium-star wind mass loss and gravitational-wave radiation. 

\subsubsection{A Roche-lobe overfilling system}
\label{ngc4214:subsubsec:rlof}

For the purpose of this work we will rely heavily on the work of Dewi et al. (2002) which explored mass transfer from helium stars when they are undergoing helium shell burning (the so-called case BB). RLOF on later stages of evolution (during core carbon burning and beyond, case BC) takes place for systems in wide periods and throughout the mass-transfer phase, the orbits never shrink to the observed 3.62-hr period (Dewi \& Pols 2003). On the other hand, in systems in closer orbits, RLOF is initiated on an earlier evolutionary stage (during helium main sequence, case BA). Dewi et al. (2002) found that dynamically stable case BA mass transfer only occurs from helium stars less massive than $\sim$ 3~\msun, and in all cases the periods never increase to reach 3.62~hr (Dewi et al. 2002). Therefore, we conclude that at the onset of RLOF, the helium star must be beyond the helium main sequence stage, but has not started the core carbon burning, i.e.\ it is undergoing a case BB mass transfer.

\begin{figure}
\centerline{\includegraphics[width=70mm, angle=270]{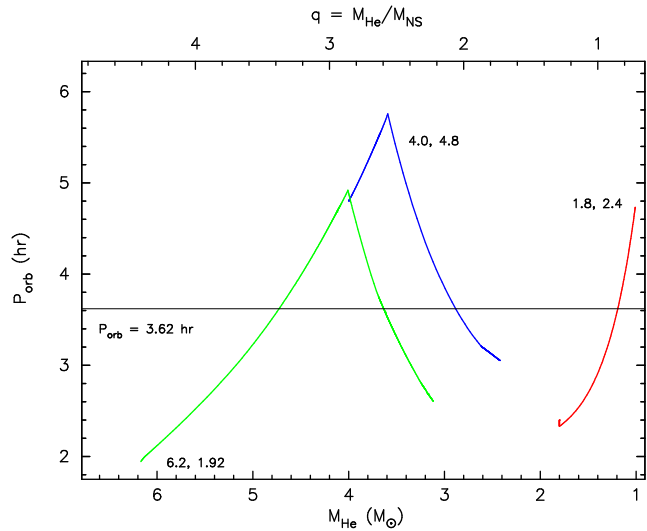}}
\caption{The evolution of the donor mass and the orbital period of helium star-neutron star binaries. The horizontal line marks the 3.62-hr orbital period of CXOU~J121538.2+361921. The pair of numbers present the combinations of initial helium-star mass (in ~\msun) and orbital period (in hours) of the plotted calculations. A scale for the mass ratio of the system is shown on the top of the figure. Except in the model of 1.8~\msun, the reverse in period marks the onset of the mass-transfer phase.}
\label{ngc4214:fig:mass-period}
\end{figure}

In the following discussion, the term {\em donor} is referred to the current mass, and {\em helium star} to the initial mass. Fig.~\ref{ngc4214:fig:mass-period} shows the evolution of the mass and orbital period for a few models, taken from the result of Dewi et al. (2002). The horizontal line presents the period of CXOU~J121538.2+361921. For each plotted model, the pair of numbers yields the initial helium-star mass in solar mass and the orbital period in hours. Thus, the red line presents the evolution of a HeS-NS system with an initial mass of 1.8~\msun\ and an orbital period of 1.92~hr. 

Dewi et al. (2002) applied the stellar wind mass loss by Hamann, Sch\"{o}nberner \& Heber (1982) and Hamann, Koesterke \& Wessolowski (1995) -- multiplied by a factor of 0.5 to provide a better fit with the most recent observed mass-loss rate of Wolf-Rayet stars (e.g., Nugis \& Lamers 2000). Before RLOF, the orbit expands due to stellar wind mass loss. During the mass-transfer phase, the period decreases. Hence, the peak in the period marks the onset of RLOF. The orbit shrinks until the period reaches a minimum when $q = 1.39$ and then expands. This critical $q$, implying a critical donor mass of 1.95~\msun, is obtained in a non-conservative mass-transfer phase where the neutron star ejects almost all the transferred matter (Dewi et al. 2002).

Helium stars more massive than 4~\msun\ experience a strong stellar wind mass loss (Pols \& Dewi 2002); the masses are reduced to around 4~\msun\ just at the onset of RLOF. An additional mass loss during the mass-transfer phase decreases the helium star masses further. We found that only massive helium stars in the initial periods of 1.9 -- 2.4 hours can evolve into the observed 3.62-hr period during mass-transfer phase. Interestingly, we found that the donor mass at the 3.62~hr period never exceeds $\sim$ 3.6~\msun\ (shown by the green line). This provides a maximum mass ratio of 2.6. At the end of the calculations, the core mass is slightly more massive than 2~\msun, and hence the donors never reach the critical mass of 1.95~\msun\ and the orbits never shrink to the minimum period.

We find that a donor mass of $\sim$ 2 -- 3~\msun\ with a 3.62-hr period is attainable from lower initial helium-star masses, i.e.\ in the range of $\sim$ 2.4 -- 4~\msun, with an initial period of around 4 -- 5~hr. The blue line in Fig.~\ref{ngc4214:fig:mass-period} shows an initially 4~\msun\ donor which evolves into the observed period when the donor mass is $\sim$ 2.8~\msun, yielding a mass ratio of 2.

In the 1.8-\msun\ model (red line), due to the weak stellar wind, the orbit prior to RLOF is dominated by gravitational-wave radiation which decreases the period. At the onset of RLOF, the helium-star mass is already lower than the critical mass for period increase, such that the orbit always expands throughout the mass-transfer phase. In the course of its evolution, the system reaches a period of 3.62~hr. However, at this point the mass of the donor is $\sim$ 1.2~\msun, i.e.\ slightly lower than the allowed mass-ratio constraint. Such a low mass ratio is allowed only if the minimum eclipse half angle is less than 22$^{o}$ (fig.~5 of Ghosh et al. 2006). However, this implies that the system is not the progenitor of a DNS because the donor mass is lower than the Chandrasekhar mass.

Ghosh et al. (2006) showed that RLOF on the terminal-age helium-star main sequence (case BA) will require a donor mass of 4.1~\msun\ for $q=10$, implying a 0.41-\msun\ compact object. Since it is more likely that the mass-transfer phase occurs beyond the helium-star main sequence, a lower-mass helium star and/or a lower mass ratio is expected (see  Sect.~\ref{ngc4214:subsec:source}). This implies that a low-mass compact accretor, i.e.\ a neutron star, is more likely.

\subsubsection{A wind-accreting system}
\label{ngc4214:subsubsec:wind}
 
Ghosh et al. (2006) have discussed the option that the system is underfilling its Roche lobe. In their illustration, if the donor radius is 75 per cent of its Roche radius and the mass ratio is 1, then the system can be explained by a donor in the range of 7~\msun\ (if it is on the zero-age helium-star main sequence) to 0.9~\msun\ (on the terminal-age helium burning). In this situation, the X-ray luminosity is generated by wind accretion from the donor star. Assuming a Bondi-Hoyle accretion (Bondi \& Hoyle 1944), one could obtain the wind-accretion rate by applying, e.g., eq.~6 of Hurley, Tout \& Pols (2002). For a 4~\msun\ helium star in a 6-hr orbit, we find that the neutron star can accrete 9 per cent of the wind mass-loss rate. This implies that to reach the required mass-accretion rate of \xpn{6}{-8}~\rate\ (as derived from the X-ray luminosity of CXOU~J121538.2+361921), a wind mass-loss rate higher than \xpn{7}{-7}~\rate\ is required from the donor.

On the helium main sequence, only helium stars more massive than $\sim$ 4.5~\msun\ release wind mass loss with a rate of the order of $10^{-6}$~\rate. The same rate can be achieved by lower-mass helium stars, i.e.\ more massive than $\sim$ 2.6~\msun,  on the stage of helium shell burning -- as wind mass loss on this stage is generally stronger than on the main sequence. In fig.~4 of Ghosh et al. 2006, the allowed area for a wind-accreting system on the helium shell burning stage is limited below the dashed red line and above the dashed blue line. Our additional constraint that the donor must be above 2.5~\msun\ leaves a very narrow parameter space. This suggests that if CXOU~J121538.2+361921 is a wind-accreting system, then it is more likely that the donor is more massive than $\sim$ 4.5~\msun\ and is on the helium main sequence. However, as a dense stellar wind was not detected (Ghosh et al. 2006),  a Roche-lobe overfilling is more likely to explain this system.

In the case of wind-accreting system, the observed orbital period can be reached by a HeS-NS binary in initially relatively short periods, in which the orbits expand due to wind mass loss. This situation is illustrated as the green line in Fig.~\ref{ngc4214:fig:mass-period}. When the system reaches the orbital period of 3.62~hr, the donor mass is $\sim$ 4.7~\msun, implying $q = 3.3$.

\section{Conclusions}
\label{ngc4214:sec:conclusions}

CXOU~J121538.2+361921 is a 3.62~hr X-ray binary with an X-ray luminosity up to \xpn{0.7}{39}~$\mathrm{erg \, s^{-1}}$. A system of a slightly evolved helium star and a neutron star or a very low mass black hole companion has been proposed (Ghosh et al. 2006). In this work we investigate this suggestion by means of detailed evolutionary calculations of a helium star with a neutron-star companion.

The X-ray luminosity can be explained by a HeS-NS binary in a state of mass transfer with a super-Eddington accretion rate, in which RLOF is initiated on the helium shell burning stage. The lack of apparent stellar wind implies that the donor is a low-mass helium star. We suggest that the compact accretor in the system is a neutron star. This puts CXOU~J121538.2+361921 as a stronger candidate for the progenitor of a DNS than Cyg X-3 since the latter probably harbours a black hole and/or a massive helium star. We find that if the X-ray binary is indeed a HeS-NS system, the observed orbital period can be reproduced from a donor of $\sim$ 2.2 -- 3.6~\msun, providing a maximum mass ratio of 2.6. A better estimate on the inclination angle of the system would improve the constrain on mass ratio (see fig.~5 of Ghosh et al. 2006), and hence on the donor mass. If further observations reveal a mass ratio less than $\sim$ 2.6, this could confirm that CXOU~J121538.2+361921 is a HeS-NS binary, and hence a double neutron star in the making.

\section*{Acknowledgments}
This work was funded by PPARC through the IoA Theory Rolling Grant. It is a pleasure to acknowledge Saul Rappaport and Philipp Podsiadlowski for their suggestion to initiate this study, and Giovanni Miniutti and Rob Fender for discussions.

\label{lastpage}

\end{document}